\documentclass[
twocolumn,
reprint,
superscriptaddress,
amsmath,
amssymb,
aps,
prl,
]{revtex4-1}
 
\usepackage[latin1]{inputenc}
\usepackage{amsmath}
\usepackage{amsfonts}
\usepackage{hyperref}
\usepackage{amssymb}
\usepackage{graphicx}
\usepackage{hyperref}
\usepackage{colortbl}
\usepackage{subfigure}
\usepackage{svg}

\usepackage[T1]{fontenc}
\usepackage{mathptmx}
\usepackage{bbold}
\usepackage{scalerel}
\usepackage{xcolor}

\usepackage{mathtools}

\newcommand*\ch[1]{\ensuremath{\mathrm{#1}}}
\newcommand{\hyref}[1]{\hyperref[#1]{\ref{#1}}}
\newcommand{\dd}{\mathrm{d}}

\newcommand{\orange}[1]

\begin{document}

\title{Iontronic neuromorphic signalling with conical microfluidic memristors}
\author{T. M. Kamsma}
\affiliation{Institute for Theoretical Physics, Utrecht University,  Princetonplein 5, 3584 CC Utrecht, The Netherlands}
\affiliation{Mathematical Institute, Utrecht University, Budapestlaan 6, 3584 CD Utrecht, The Netherlands}
\author{W. Q. Boon}
\affiliation{Institute for Theoretical Physics, Utrecht University,  Princetonplein 5, 3584 CC Utrecht, The Netherlands}
\author{T. ter Rele}
\affiliation{Institute for Theoretical Physics, Utrecht University,  Princetonplein 5, 3584 CC Utrecht, The Netherlands}
\affiliation{Soft Condensed Matter, Debye Institute for Nanomaterials Science, Utrecht University,  Princetonplein 1, 3584 CC Utrecht, The Netherlands}
\author{C. Spitoni}
\affiliation{Mathematical Institute, Utrecht University, Budapestlaan 6, 3584 CD Utrecht, The Netherlands}
\author{R. van Roij}
\affiliation{Institute for Theoretical Physics, Utrecht University,  Princetonplein 5, 3584 CC Utrecht, The Netherlands}

\date{\today}

\begin{abstract}
Experiments have shown that the conductance of conical channels, filled with an aqueous electrolyte, can strongly depend on the history of the applied voltage. These channels hence have a memory and are promising elements in brain-inspired (iontronic) circuits. We show here that the memory of such channels stems from transient concentration polarization over the ionic diffusion time. We derive an analytic approximation for these dynamics which shows good agreement with full finite-element calculations. Using our analytic approximation, we propose an experimentally realisable Hodgkin-Huxley iontronic circuit where micrometer cones take on the role of sodium and potassium channels. Our proposed circuit exhibits key features of neuronal communication such as all-or-none action potentials upon a pulse stimulus and a spike train upon a sustained stimulus.
\end{abstract}

\maketitle

Transport phenomena of charged species through channels in the nanometre and micrometre regime play a key role in a plethora of applications \cite{knust2014electrochemical, surwade2015water,lai2015desalination,van2007power,siria2013giant,xiao2019ion,saleh2003artificial,shi2017nanopore, vlassiouk2009biosensing,de2012nanochannels,zhang2016fundamental,siwy2023nanopores}. An exciting emerging direction of research is that of iontronics, the use of ion transport for signalling \cite{han2022iontronics}, which holds the promise of interfacing with biological systems \cite{han2022iontronics,yang2019cavity,hu2019ultrasensitive,shchanikov2021designing} and processing information via multiple signal carriers and chemical regulation \cite{noy2023nanofluidic,li2020synaptic}. In particular conical channels have garnered significant interest for such applications \cite{yang2019cavity,hu2019ultrasensitive,jubin2018dramatic, hou2011biomimetic, ghosal2019solid, bush2020chemical, liu2012surface}, as they exhibit current rectification thereby acting as ionic diodes \cite{wei1997current, boon2021nonlinear, white2008ion, jubin2018dramatic, vlassiouk2009biosensing}. This has been extensively studied experimentally \cite{cheng2007rectified, siwy2006ion, bush2020chemical, jubin2018dramatic, siwy2002rectification,fulinski2005transport,siwy2005asymmetric} as well as numerically \cite{duleba2022effect, lan2016voltage, vlassiouk2008nanofluidic, liu2012surface, kubeil2011role} and several analytic descriptions are available \cite{boon2021nonlinear, dal2019confinement, poggioli2019beyond,uematsu2022analytic}. Due to this interest cones are now comparatively easy to fabricate  \cite{kovarik2009effect, lin2018voltage, siwy2003preparation, siwy2002fabrication}. Recently, it has been observed that the conductance of cones exhibits hysteresis when driven by an alternating potential \cite{wang2012transmembrane,  li2015history, wang2014physical,  wang2016dynamics, wang2017correlation, sheng2017transporting, brown2020deconvolution, brown2022selective, brown2021higher, wang2018hysteresis,  ramirez2021negative}, and hence they are memristors (resistors with memory) \cite{chua1971memristor,strukov2008missing,chua2014if,caravelli2018memristors}. Various explanations of this effect have been explored \cite{wang2012transmembrane, li2015history, wang2014physical, wang2017correlation,wang2016dynamics, sheng2017transporting}, the most recent hypothesis being dynamic concentration polarization \cite{wang2018hysteresis, brown2021higher, brown2022selective, ramirez2021negative}, which we confirm below. Memristors in general are essential components for neuromorphic (brain-inspired) circuits, since much of their dynamics is analogous to the synapses that connect neurons \cite{chua2013memristor,van2018organic,keene2021neuromorphic, chicca2020recipe,christensen20222022} and to the ion channels responsible for electric signaling within neurons \cite{sah2014brains,chua2013memristor}. The popularity of memristors and neuromorphic circuits has drastically increased \cite{schuman2017survey,venkatesan2022brain,zhu2020comprehensive} due to the prospects of energy-efficient computers \cite{mehonic2022brain,sangwan2020neuromorphic,caravelli2021global} and bio-compatibility \cite{keene2020biohybrid,harikesh2022organic,krauhausen2021organic,marasco2021neurorobotic,van2018organic,yuan2021organic}. Solid-state devices received the majority of attention \cite{schuman2017survey,venkatesan2022brain,sangwan2020neuromorphic,zhu2020comprehensive}, while the brain in contrast relies on ion transport in an aqueous medium \cite{fundNeuroE,bean2007action}. Iontronic circuits, based on the same signalling medium as the brain, sparked recent interest as a promising platform for a new generation of (neuromorphic) computing devices \cite{noy2023nanofluidic,xiong2023neuromorphic,robin2023long,xie2022perspective,kim2023liquid}. However, the development of neuromorphic iontronic devices is still in its infancy \cite{han2022iontronics,xie2022perspective} and a deeper understanding of underlying mechanisms and possible end-uses is needed \cite{xie2022perspective,kim2023liquid}.

\begin{figure}[ht]
	\centering
	\includegraphics[width=0.42\textwidth]{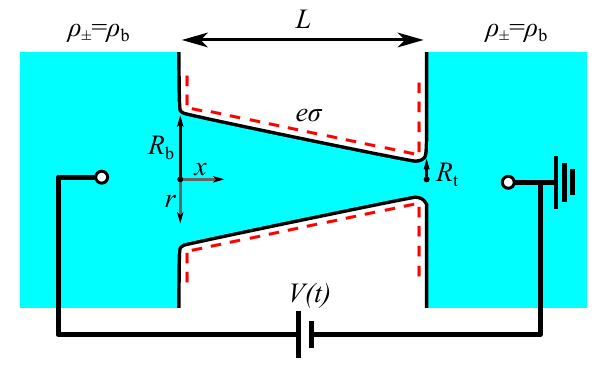}\vspace{-0.4 cm}
	\caption{Schematic representation of an azimuthally symmetric conical channel of length $L$, base radius $R_{\mathrm{b}}$, and tip radius $R_{\mathrm{t}}<R_{\mathrm{b}}$, connecting two bulk reservoirs of a 1:1 aqueous electrolyte, with bulk concentrations $\rho_{\mathrm{b}}$. The channel wall carries a surface charge density $e\sigma$. An AC electric potential drop $V(t)$ over the channel drives an ionic charge current $I(V(t),t)=g(V(t),t)V(t)$ with $g(V(t),t)$ the channel conductance. The AC potential causes transient concentration polarisation resulting in a volatile conductance memory.}
	\label{fig:setup}
	\vspace{-0.25 cm}
\end{figure}
In this Letter we propose a neuromorphic iontronic circuit where all-or-none action potentials \cite{lucas1909all,bean2007action,fundNeuroE} and neuronal spiking \cite{fundNeuroTrain,cymbalyuk2002bursting,marder2002cellular,sherman2001tonic,bean2007action} are obtained through micrometer cones filled with aqueous electrolyte. First we solve for the dynamic competition between ionic diffusion, conduction and advection, using the Poisson-Nernst-Planck-Stokes equations and the steady-state results in Ref.~\cite{boon2021nonlinear}. We obtain a differential equation with no free parameters describing transient concentration polarization and we surprisingly find that the inhomogeneous conduction responsible for concentration polarization retains memory over slow (diffusive) timescales. Using this result we construct an experimentally accessible iontronic circuit of several cones, batteries, and a capacitor and show that the time-dependent voltage over this circuit exhibits multiple key features of neuronal communication. While neuronal behaviour in the form of an emerging voltage spike train has been reported in a simulated iontronic circuit containing quasi two-dimensional nanochannels that connect aqueous electrolytes \cite{robin2021principles}, the defining all-or-none law of action potentials \cite{lucas1909all,bean2007action,fundNeuroE}, a phenomenon considered to be a requirement for artificial neurons \cite{zhu2020comprehensive}, has not yet been reported. Furthermore the circuit in Ref.~\cite{robin2021principles} requires comparatively difficult to fabricate two-dimensional channels and Nernst potentials that do not directly affect the circuit voltage, complicating experimental accessibility. In this Letter we overcome these issues. As conical pores are a well-known iontronic model system, we believe our proposed neuromorphic circuit is within experimental reach. 

We first consider a single conical channel, schematically illustrated in Fig.~\ref{fig:setup}, that connects two bulk reservoirs of an incompressible aqueous 1:1 electrolyte with viscosity $\eta=1.01\text{ mPa}\cdot\text{s}$, mass density $\rho_{\mathrm{m}}=10^3\text{ kg}\cdot\text{m}^{-3}$ and electric permittivity $\epsilon=0.71\text{ nF}\cdot\text{m}^{-1}$, containing ions with diffusion coefficients $D_{\pm}=D=1.75\text{ }\mu\text{m}^2\text{ms}^{-1}$ and charge $\pm e$ with $e$ the proton charge. At the far side of both reservoirs we impose a fixed pressure $P=P_0$ and fixed ion concentrations $\rho_{\pm}=\rho_{\mathrm{b}}=0.1\text{ mM}$. The reservoirs are connected by an azimuthally symmetric conical channel with base radius $R_{\mathrm{b}}=200\text{ nm}$ at $x=0$ and tip radius $R_{\mathrm{t}}=R_{\mathrm{b}}-\Delta R=50\text{ nm}$ at $x=L\gg R_{\mathrm{b}}$, the central axis being at radial coordinate $r=0$. Unless otherwise stated the channel has length $L=10\text{ }\mu\text{m}$, hence the geometry is similar to cones in Ref.~\cite{kovarik2009effect}. The channel radius is described by $R(x)=R_{\mathrm{b}}-x\Delta R/L$ for $x\in\left[0,L\right]$. We assume a uniform surface charge density $e\sigma=-0.0015\;e\text{nm}^{-2}$ on the channel walls, resulting in a surface potential $\psi_0\approx -10\text{ mV}$ (typical for PMMA \cite{kirby2004zeta}) and an electric double layer that screens the surface charge with Debye length $\lambda_{\mathrm{D}}\approx30\text{ nm}$. On the far side of the reservoir connected to the base we impose an electric potential $V(t)$, while the far side of the other reservoir is grounded, which leads to an electric potential profile $\Psi(x,r,t)$, an electro-osmotic fluid flow with velocity field $\mathbf{u}(x,r,t)$ and ionic fluxes $\mathbf{j}_{\pm}(x,r,t)$ with $\mathbf{j}_{+}-\mathbf{j}_{-}$ the charge flux. A relatively low surface potential $\psi_0$ ensures a weak electro-osmotic flow $Q(V)$, allowing conductance tuning over a wider voltage range \cite{aarts2022ion, boon2021nonlinear, brown2020deconvolution}. We have $Q(V)/V=-\pi R_{\mathrm{t}}R_{\mathrm{b}}\epsilon\psi_0/(\eta L)\approx22.7\text{ }\mu\ch{m}^3\text{s}^{-1}\text{V}^{-1}$ for our standard parameter set \cite{boon2021nonlinear}.

Transport through the conical channel is described by the Poisson-Nernst-Planck-Stokes (PNPS) Eqs.~(\ref{eq:poisson})-(\ref{eq:stokes}) given by 
\begin{gather}
	\nabla^2\Psi=-\frac{e}{\epsilon}(\rho_+-\rho_-),\label{eq:poisson}\\
	\dfrac{\partial\rho_{\pm}}{\partial t}+\nabla\cdot\mathbf{j}_{\pm}=0,\label{eq:ce}\\
 \mathbf{j}_{\pm}=-D_{\pm}\left(\nabla\rho_{\pm}\pm\rho_{\pm}\frac{e\nabla \Psi}{k_{\mathrm{B}}T}\right)+\mathbf{u}\rho_{\pm},\label{eq:NP}\\
	\rho_{\mathrm{m}}\dfrac{\partial\mathbf{u}}{\partial t}=\eta\nabla^2\mathbf{u}-\nabla P-e(\rho_+-\rho_-)\nabla \Psi;\qquad\nabla\cdot\mathbf{u}=0.\label{eq:stokes}
\end{gather}
Here electrostatics is accounted for by the Poisson Eq.~(\ref{eq:poisson}), the conservation of ions by the continuity Eq.~(\ref{eq:ce}), the combination of Fickian diffusion, Ohmic conduction, and Stokesian advection by the Nernst-Planck Eq.~(\ref{eq:NP}), and finally the force balance on the (incompressible) fluid by the Stokes Eq.~(\ref{eq:stokes}). This system is closed upon imposing no-slip and blocking boundary conditions on the channel wall, $\mathbf{u}=0$ and $\mathbf{n}\cdot\mathbf{j}_{\pm}=0$, respectively, together with  Gauss' law $\mathbf{n}\cdot\nabla\Psi=-e\sigma/\epsilon$, with $\mathbf{n}$ the wall's inward normal vector.

When an electric potential $V(t)$ is applied, the ionic concentrations $\rho_{\pm}(x,r,t)$ will deviate from their equilibrium profiles, thereby changing the channel conductance  \cite{boon2021nonlinear}. In Ref.~\cite{boon2021nonlinear} the stationary state version of the PNPS equations (\ref{eq:poisson})-(\ref{eq:stokes}) is solved for a static potential $V$, which gives rise to the stationary charge current $I=g_{\infty}(V)  V$, where the static conductivity $g_\infty(V)$ was found to be governed by
the (voltage-dependent) radially averaged salt concentration $\overline{\rho}_{\mathrm{s}}(x,V)=2\int_{0}^{R(x)}r\rho_{\mathrm{s}}(x,V,r)\dd r/R(x)^2$, with $\rho_{\mathrm{s}}=\rho_{+}+\rho_{-}$ (see Supplemental Material for full expression).
For small potentials $e|V|/k_{\rm{B}}T\ll |w| (R_{\rm{b}}/R_{\rm{t}})$, with $w=eD\eta/(k_{\rm{B}}T\epsilon \psi_0)\simeq-9.5$ the ratio of ionic to electro-osmotic mobility \cite{aarts2022ion}, the pore concentration equals the bulk concentration $\overline{\rho}_{\mathrm{s}}(x,V)=2\rho_{\mathrm{b}}$, yielding $g_{\infty}(V)=g_0$, with the Ohmic cone conductance $g_0=(\pi R_{\mathrm{t}} R_{\mathrm{b}}/L)(2\rho_{\rm{b}}e^2D/k_{\mathrm{B}}T)$, and the resulting current follows Ohm's law $I=g_0 V$ \cite{boon2021nonlinear}. For large static potential drops the cone exhibits diodic behaviour due to concentration polarisation, with the cone conductance determined by the salt concentration profiles according to 
\begin{equation}\label{eq:rhos}
    \begin{aligned}
    \frac{g_{\infty}(V)}{g_0}=&\int_0^L \overline{\rho}_{\mathrm{s}}(x,V) \dd x/(2\rho_{\rm{b}}L)\\
    =&1+\Delta g\int_0^{L}\left[\frac{x}{L}\frac{R_{\mathrm{t}}}{R(x)}-\frac{e^{\text{Pe}(V)\frac{x}{L}\frac{ R_{\mathrm{t}}^2}{R_{\mathrm{b}}R(x)}}-1}{e^{\text{Pe}(V)\frac{R_{\mathrm{t}}}{R_{\mathrm{b}}}}-1}\right]\dd x/L,
    \end{aligned}
\end{equation}
where an approximation is made compared to the more accurate dependence on $L/\int_{0}^{L}(\overline{\rho}_{\mathrm{s}}(x,V))^{-1} \dd x$ to reduce computational complexity \cite{boon2021nonlinear}. The static conductance $g_{\infty}(V)$ depends on $V$ through the P\'{e}clet number at the narrow end $\text{Pe}(V)=Q(V)L/(D\pi R_{\ch{t}}^2)=-(eV/k_{\rm{B}}T)(R_{\mathrm{b}}/R_{\mathrm{t}})w^{-1}$, where for our standard parameters $\text{Pe}(V)/V\approx16.5 \text{ V}^{-1}$, and $\Delta g\equiv-2w(\Delta R/R_{\ch{b}})\text{Du}\approx-3.6$ with the tip Dukhin number $\text{Du}=\sigma/(2\rho_{\mathrm{b}}R_{\ch{t}})\approx-0.25$. In our case of $\sigma<0$, $V>0$ depletes the channel of ions such that $g_{\infty}(V)/g_0<1$ whereas $ V<0$ results in ion accumulation such that $g_{\infty}(V)/g_0>1$, which is responsible for the static diode behaviour of the cone \cite{boon2021nonlinear}. 

\begin{figure}[ht]
\centering
     \includegraphics[width=1\textwidth]{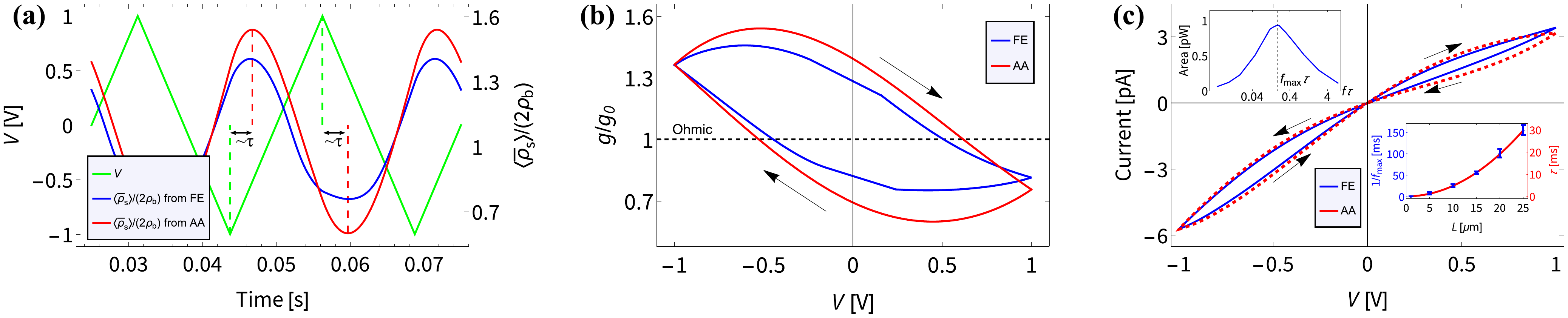}\vspace{-0.4 cm}
        \caption{Comparisons of finite-element calculations (FE, blue) of the full PNPS equations (\ref{eq:poisson})-(\ref{eq:stokes})  and our analytic approximation Eqs.~(\ref{eq:rhos})-(\ref{eq:I}) (AA, red) when a periodic triangle potential $V(t)$ (green in \textbf{(a)}) with amplitudes $\pm 1 \text{ V}$ and frequency $f=40\text{ Hz}$ is applied. In \textbf{(a)} we show two periods of the time-dependent laterally averaged salt concentration  $\langle\overline{\rho}_{\mathrm{s}}\rangle(V(t),t)\equiv\int_{0}^{L}\overline{\rho}_{\mathrm{s}}(x,V(t),t)\dd x/L$, in \textbf{(b)} the corresponding conductance-voltage diagram as per Eqs.~(\ref{eq:rhos}) and (\ref{eq:deqrho}), and in \textbf{(c)} the corresponding current-voltage diagram with a clear pinched hysteresis loop. The left inset of \textbf{(c)} shows the dependence of the enclosed area in I-V hysteresis loop on the (dimensionless) triangle potential frequency $f\tau$ with a maximum at $f_{\mathrm{max}}\tau\approx 0.19$ for $L=10\;\mu\ch{m}$. The right inset of \textbf{(c)} shows the dependence of $1/f_{\ch{max}}$ (blue) and of the characteristic time $\tau$ from Eq.~(\ref{eq:ts}) (red) on the channel length $L$, satisfying $f_{\mathrm{max}}\tau\approx 0.19$ for all $L$ considered.}
        \label{fig:theorycompare}
\end{figure}

Voltage driven accumulation or depletion of ions from the pore is not instantaneous, as it takes time for the ions to move into or out of the channel. In the Supplemental Material (SM) \cite{SM} we derive from the PNPS equations (\ref{eq:poisson})-(\ref{eq:stokes}) that the typical timescale for this process is not given by an RC-like time as suggested in Ref.~\cite{wang2016dynamics}, but rather by a diffusion-like time
\begin{align}
    \tau=\frac{L^2}{12D},\label{eq:ts} 
\end{align}
which for our standard parameter set yields $\tau=4.8\text{ ms}$. To obtain an analytic approximation for the time-dependent conductance $g(V(t),t)$ we assume a single exponential relaxation of the salt-concentration with timescale $\tau$ towards the steady-state concentration profile. This natural approach has been successfully applied to investigate memristor dynamics before \cite{robin2023long,markin2014analytical} and is verified for conical channels in Fig.~\ref{fig:theorycompare}. Using Eq.~(\ref{eq:rhos}), this approach yields the following expressions for the time-dependent conductance $g(V(t),t)$ and current $I(V(t),t)$
\begin{equation}
	\dfrac{\partial g(V(t),t)}{\partial t}=\frac{g_\infty( V(t))-g(V(t),t)}{\tau},\label{eq:deqrho}
\end{equation}
\begin{equation}
	I(V(t),t)=g(V(t),t)V(t).\label{eq:I}
\end{equation}
\onecolumngrid\
\twocolumngrid\
Eq.~(\ref{eq:deqrho}) shows that the conductance $g(V(t),t)$ depends on the entire timetrace of the potential $V(t)$. This conductance memory is the key feature of a memristor \cite{caravelli2018memristors}, in fact Eqs.~(\ref{eq:deqrho}) and (\ref{eq:I}) indeed fit the mathematical definition of a generic voltage-driven memristor \cite{chua2014if,caravelli2018memristors,chua2013memristor}. From now on we will refer to Eqs.~(\ref{eq:rhos})-(\ref{eq:I}) as the approximate analytical (AA) solution of the PNPS equations \cite{boon2021nonlinear}. To verify the AA we also numerically solve the full PNPS equations (\ref{eq:poisson})-(\ref{eq:stokes}) in the geometry of a conical channel using the finite-element (FE) analysis package COMSOL \cite{multiphysics1998introduction,pryor2009multiphysics}.

In Fig.~\ref{fig:theorycompare}(a) we compare the time-dependent laterally averaged salt concentrations $\langle\overline{\rho}_{\mathrm{s}}\rangle(V(t),t)\equiv\int_{0}^{L}\overline{\rho}_{\mathrm{s}}(x,V(t),t)\dd x/L$ from FE calculations (blue) and the AA (red), which for the AA is equivalent to $g(V(t),t)/g_0$. In both cases $V(t)$ is a periodic triangle potential (green) with amplitudes $\pm 1 \text{ V}$ and a period of $25\text{ ms}$. The features of the AA and FE calculations essentially agree, not only the typical amplitude of $\langle\overline{\rho}_{\mathrm{s}}\rangle(V(t),t)$, but also the time lag $\sim\tau$ (Eq.~(\ref{eq:ts})) between $V(t)$ and $\langle\overline{\rho}_{\mathrm{s}}\rangle(V(t),t)$, as indicated by the two pairs of dashed vertical lines. This time lag results in a hysteretic conductance-voltage diagram, shown in Fig.~\ref{fig:theorycompare}(b). Here the AA and FE calculations agree with each other again, verifying the proposed relation of $\langle\overline{\rho}_{\mathrm{s}}\rangle(V(t),t)$ with the conductance $g(V(t),t)$.

Translating the results of Fig.~\ref{fig:theorycompare}(b) to the current with Eq.~(\ref{eq:I}) we obtain the current-voltage (I-V) plot of Fig.~\ref{fig:theorycompare}(c), which shows the memristor hallmark of a pinched hysteresis loop \cite{chua2014if}. We again find agreement between the AA and FE calculations and the I-V loop resembles previously reported experimental results from comparable systems \cite{brown2020deconvolution,li2015history,wang2012transmembrane,wang2014physical,wang2018hysteresis}. The shape of the hysteresis loop depends on the frequency $f$ of the applied triangle potential $V(t)$. The enclosed area inside the loop shrinks to 0 for $f\tau\ll1$ and $f\tau\gg1$ and shows a maximum at $f_{\rm{max}}\tau\approx 0.19$ for the standard parameter set, as shown in the left inset of Fig.~\ref{fig:theorycompare}(c). In the right inset of Fig.~\ref{fig:theorycompare}(c) we see that the one-to-one relation $f_{\rm{max}}\tau\approx 0.19$ also holds for various lengths $L$, further supporting the validity of Eq.~(\ref{eq:ts}). Excitingly, the quadratic dependence of $\tau$ on the channel length $L$ offers strong control over the channel memory retention time, a desirable trait \cite{chicca2020recipe}.

\begin{figure}[ht]
	\centering
	\includegraphics[width=0.95\textwidth]{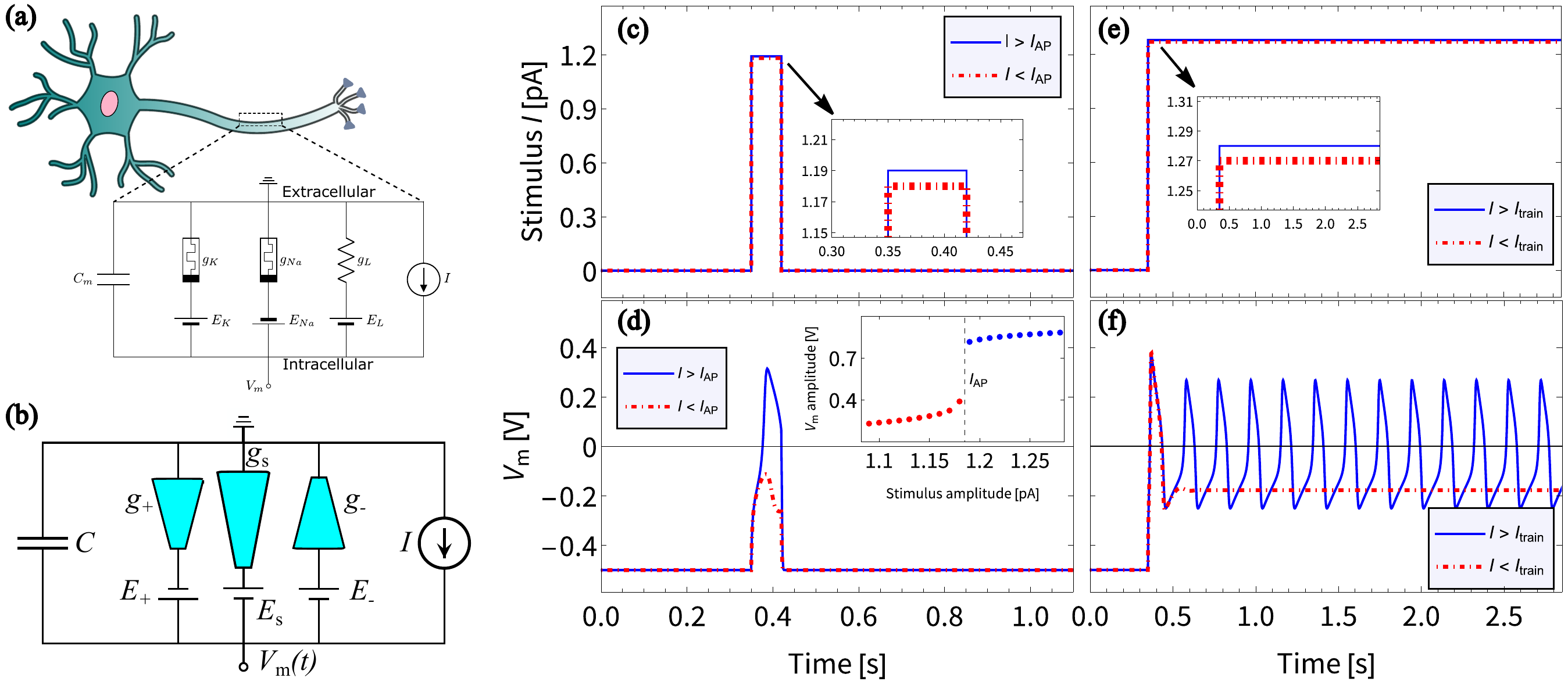}
	\caption{\textbf{(a)} The Hodgkin-Huxley circuit model with $\mathrm{Na}^{+}$, $\mathrm{K}^{+}$ and leak channels, a capacitor and batteries corresponding to the $\mathrm{Na}^{+}$ and $\mathrm{K}^{+}$ Nernst potentials \cite{hodgkin1952quantitative}. \textbf{(b)} Schematic representation of our proposed circuit containing three oriented conical channels, connected in series to individual batteries and in parallel to a capacitor. The electric potential difference $V_{\ch{m}}(t)$ over the capacitor can be driven by an imposed stimulus current $I(t)$. \textbf{(c)} The imposed subcritical (red) and supercritical (blue) current pulse $I(t)$, and \textbf{(d)} the resulting $V_{\mathrm{m}}(t)$ from Eq.~(\hyref{eq:HH_3_2}), displaying an all-or-none action potential, as can be seen by the jump in spike amplitude around $I_{\mathrm{AP}}$ as shown in the inset. \textbf{(e)} The imposed subcritical (red) and supercritical (blue) sustained currents $I(t)$, and \textbf{(f)} the resulting $V_{\mathrm{m}}(t)$, where a spike train emerges for $I(t)>I_{\ch{train}}$.}
	\label{fig:neurofigs}
\end{figure}

Having derived Eqs.~(\ref{eq:ts})-(\ref{eq:I}) for the memristive effect in a single conical channel, we now turn our attention to modelling a brain-inspired iontronic circuit. Electric signalling within a neuron is facilitated by an action potential (AP), a propagating voltage spike over the cell membrane \cite{fundNeuroE}. APs obey the all-or-none law, i.e.\ an AP either fails to initiate upon a subcritical stimulus or completely occurs for a supercritical stimulus, with no gradual transition in between \cite{lucas1909all,bean2007action,fundNeuroE}, and can be sequentially generated, resulting in a spike train \cite{fundNeuroTrain,cymbalyuk2002bursting,marder2002cellular,sherman2001tonic,bean2007action}. These neuronal features of electric activity over the membrane have been successfully modelled by an equivalent circuit as in Fig.~\ref{fig:neurofigs}(a), first quantitatively characterised by Hodgkin and Huxley \cite{hodgkin1952quantitative}, which has formed an extensively used basis to mathematically model neuronal signalling  \cite{rall2011core,fitzhugh1973dimensional,rall1962theory,halter1991distributed,hay2011models,hines1997neuron,kole2008action}. Interestingly, the mathematical descriptions of the biological potassium and sodium channel conductances $g_{\mathrm{K}}$ and $g_{\mathrm{Na}}$ in the Hodgkin-Huxley (HH) model were later identified to be descriptions of memristors \cite{sah2014brains}. Therefore we expect similar spontaneous neuronal features by assembling conical channels in a HH-like circuit, where the micrometer channels take on the role of the potassium and sodium channels.

Inspired by HH circuits we present the circuit shown in Fig.~\ref{fig:neurofigs}(b). This circuit consists of a capacitor with capacitance $C=5\text{ }\text{fF}$ (corresponding to the  typical capacitance of a biological neuronal membrane of area $\sim 1\;\mu\text{m}^2$ \cite{gentet2000direct}), connected in parallel with three oriented conical channels, with conductances $g_{+}$, $g_{-}$ and $g_{\mathrm{s}}$ and lengths $L_{\pm}=1\text{ }\mu\text{m}$ and $L_{\mathrm{s}}=25\text{ }\mu\text{m}$. As per Eq.~(\ref{eq:ts}), the timescales $\tau_{\pm}\approx 0.048$ ms of the two fast channels are identical, while the timescale $\tau_{\mathrm{s}}\approx 30 \text{ ms } \gg\tau_{\pm}$ is much slower. The conical channels are connected in series to batteries with potentials $E_{\pm}=\pm0.975\text{ V}$ for the two fast channels and $E_{\mathrm{s}}=-0.5\text{ V}$ for the slow channel. The imposed stimulus current $I(t)$ is the control parameter and determines whether spiking occurs. The electric potential $V_{\mathrm{m}}(t)$ over the circuit shown in Fig. \ref{fig:neurofigs}(b) is equivalent to the membrane potential over a neuronal membrane \cite{hodgkin1952quantitative}.
\onecolumngrid\
\twocolumngrid\
Invoking Kirchoff's law, the potential $V_{\mathrm{m}}(t)$ will evolve according to
\begin{align}
	C\dfrac{\dd V_{\mathrm{m}}(t)}{\dd t}=I(t)-\sum_{i\in\left\{+,-,\mathrm{s}\right\}}g_{i}(V_{i}(t),t)\left(V_{\mathrm{m}}(t)-E_{i}\right),\label{eq:HH_3_2}
\end{align}
where the conductances $g_{i}(V_{i}(t),t)$, determined by their individual laterally averaged salt concentrations as per Eq.~(\ref{eq:rhos}), each evolve according to Eq.~(\ref{eq:deqrho}), however with arguments $V_i$ of $g_{i,\infty}(V_i)$ given by $V_-(t)=V_{\mathrm{m}}(t)-E_-$, $V_+(t)=-V_{\mathrm{m}}(t)+E_+$ and $V_{\mathrm{s}}(t)=-V_{\mathrm{m}}(t)+E_{\mathrm{s}}$. The differences in signs of the potentials reflect the different orientations of the channels as depicted in Fig.~\ref{fig:neurofigs}(b). Eqs.~(\ref{eq:ts}), (\ref{eq:deqrho}) and (\ref{eq:HH_3_2}) form a closed set of equations, which we numerically solve with initial conditions $V(0)=-0.5\text{ V}$ and $g_{i}(V_{i}(0),0)=g_{0,i}$.

Fig.~\ref{fig:neurofigs}(c) shows a subcritical pulse current $I<I_{\mathrm{AP}}$ (red) and supercritical pulse current $I>I_{\mathrm{AP}}$ (blue), both of duration $70$ ms, and Fig.~\ref{fig:neurofigs}(d) shows the two resulting membrane potentials $V_{\mathrm{m}}(t)$. For the supercritical stimulus $V_{\mathrm{m}}(t)$ fully depolarises and an AP emerges, while it fails to properly depolarise for the subcritical stimulus. The inset of Fig.~\ref{fig:neurofigs}(d) shows that the amplitude of the voltage spike undergoes a sharp step increase at the pulse strength $I=I_{\mathrm{AP}}$, i.e.\ the circuit exhibits the defining all-or-none law found in biological neurons \cite{lucas1909all,bean2007action,fundNeuroE}, a feature considered to be a requirement for artificial neurons \cite{zhu2020comprehensive}.

For a slightly increased and sustained stimulus strength (as shown in Fig.~\ref{fig:neurofigs}(e)), the spike train of Fig.~\ref{fig:neurofigs}(f) emerges for a supercritical input $I>I_{\mathrm{train}}$, while just a single AP appears for a subcritical input $I<I_{\mathrm{train}}$. Spike trains are another unique feature of neuromorphic behaviour and play a vital role in neuronal communication \cite{fundNeuroTrain,cymbalyuk2002bursting,marder2002cellular,sherman2001tonic,bean2007action}. In the SM \cite{SM} we show that the frequency of the spike train can be tuned by altering the capacitance and cone lengths. In Ref.~\cite{robin2021principles} a spike train emerging from a simulated iontronic circuit containing quasi two-dimensional nanochannels was presented. However, the defining all-or-none law of APs \cite{zhu2020comprehensive,lucas1909all,bean2007action,fundNeuroE} was not reported in Ref.~\cite{robin2021principles}. Additionally, in biological neurons the battery potentials stem from Nernst potentials, which are not considered to affect the voltage-gated channel conductances, hence $\left|V_i\right|=\left|V_{\mathrm{m}}\right|$ in typical HH studies \cite{hodgkin1952quantitative,rall2011core,fitzhugh1973dimensional,rall1962theory,halter1991distributed,hay2011models,hines1997neuron,kole2008action}. However, in an experimental realisation of a microfluidic HH-like circuit with electric batteries the voltages $V_{i}(t)$ over the channels will be affected by the battery potentials $E_i$, i.e.\ $\left|V_i(t)\right|=\left|V_{\mathrm{m}}(t)-E_i\right|$. This detail is not considered in Ref.~\cite{robin2021principles} and hence it is not immediately clear how this circuit could be experimentally realized.

In summary, we derived a theoretical model with no free parameters, starting from the Poisson-Nernst-Planck-Stokes equations (\hyref{eq:poisson})-(\hyref{eq:stokes}), that explains how dynamic concentration polarization in conical pores facilitates a volatile conductance memory. Our theory agrees quantitatively with the memristive conductance of conical channels observed in finite-element calculations and we surprisingly find that the conductive memory retention process is governed by a slow diffusive timescale. By assembling multiple conical channels in an experimentally accessible iontronic Hodgkin-Huxley circuit we find emerging neuronal behaviour. The circuit exhibits all-or-none action potentials upon pulse stimulation, a fundamental requirement for artificial neurons \cite{zhu2020comprehensive}, and spike trains under a sustained stimulus, thereby displaying hallmark features of neuronal communication \cite{lucas1909all,bean2007action,fundNeuroTrain,cymbalyuk2002bursting,marder2002cellular,sherman2001tonic,bean2007action}. Our work promises to accelerate the targeted development of iontronic circuits and a more effortless scanning of possible applications thereof, beyond what is presented in this Letter.

\begin{acknowledgments}
This work is part of the D-ITP consortium, a program of the Netherlands Organisation for Scientific Research (NWO) that is funded by the Dutch Ministry of Education, Culture and Science (OCW). T.M.K.\ performed the calculations; W.Q.B.\ conceptualized the work; T.M.K.\ and W.Q.B.\ developed the theory under supervision of C.S.\ and R.v.R. All authors discussed the results and contributed to the manuscript.
\end{acknowledgments}


%

\end{document}


\title{Supplemental Material for ``Iontronic neuromorphic signalling with conical microfluidic memristors''}
	\author{T. M. Kamsma}
	\affiliation{Institute for Theoretical Physics, Utrecht University,  Princetonplein 5, 3584 CC Utrecht, The Netherlands}
	\affiliation{Mathematical Institute, Utrecht University, Budapestlaan 6, 3584 CD Utrecht, The Netherlands}
	\author{W. Q. Boon}
	\affiliation{Institute for Theoretical Physics, Utrecht University,  Princetonplein 5, 3584 CC Utrecht, The Netherlands}
	\author{T. ter Rele}
	\affiliation{Institute for Theoretical Physics, Utrecht University,  Princetonplein 5, 3584 CC Utrecht, The Netherlands}
	\affiliation{Soft Condensed Matter, Debye Institute for Nanomaterials Science, Utrecht University,  Princetonplein 1, 3584 CC Utrecht, The Netherlands}
	\author{C. Spitoni}
	\affiliation{Mathematical Institute, Utrecht University, Budapestlaan 6, 3584 CD Utrecht, The Netherlands}
	\author{R. van Roij}
	\affiliation{Institute for Theoretical Physics, Utrecht University,  Princetonplein 5, 3584 CC Utrecht, The Netherlands}
	\maketitle
	\setcounter{figure}{0}
	\renewcommand{\figurename}{Figure}
	\renewcommand{\thefigure}{S\arabic{figure}}
	\section{Channel memory timescale and range of validity}\label{sec:timescale}
	To derive the memory retention timescale $\tau$ of a conical channel from the PNPS equations, we consider two quantities, the change of total number of ions in the channel and the net salt flux. The total number of ions in the channel $N=\pi\int_0^LR^2(x)\overline{\rho}_{\mathrm{s}}(x,V)\dd x$ changes when a voltage is applied over the channel. Using results from Ref.~\cite{boon2021nonlinear} for the radially averaged steady-state salt concentration at a static voltage $V$,
	\begin{equation}\label{eq:rhosProf}\tag{S1}
		\overline{\rho}_{\mathrm{s}}(x,V)=2\rho_{\ch{b}}+2\rho_{\ch{b}}\Delta g\left[\frac{x}{L}\frac{R_{\mathrm{t}}}{R(x)}-\frac{e^{\text{Pe}(V)\frac{x}{L}\frac{ R_{\mathrm{t}}^2}{R_{\mathrm{b}}R(x)}}-1}{e^{\text{Pe}(V)\frac{R_{\mathrm{t}}}{R_{\mathrm{b}}}}-1}\right],
	\end{equation}
	we find that the change in $N$ upon a small voltage perturbation $V^{\prime}$ around $V=0$ yields
	\begin{equation}\label{eq:alpha}\tag{S2}
		\begin{aligned}
			\left.\dfrac{\partial N}{\partial V}\right|_{V=0} V^{\prime}=&\left(\pi\int_0^LR^2(x)\dfrac{\partial \overline{\rho}_{\mathrm{s}}(x,V)}{\partial V}\dd x\right)V^{\prime}\\
			=&\frac{\pi}{6}L\Delta R\frac{e\sigma}{k_BT}V^{\prime}\equiv \alpha V^{\prime},
		\end{aligned}
	\end{equation}
	where $\alpha<0$ for our parameter choice of $\sigma<0$ and $\Delta R>0$, in agreement with the enhanced (reduced) conductance of a negative (positive) potential $V^{\prime}$. 
	
	We can find the time it takes to add $\alpha V^\prime$ ions to the channel by considering the salt current $J$. For the net total influx of salt into the channel due to a change in the static electric potential we use the expression for the total salt flux, i.e.\ the $x$-component of $J$, given in Ref.~\cite{boon2021nonlinear} by
	\begin{equation}\label{eq:saltflux}
		\begin{split}
			J_x(x)=&-D\left(\pi R^2(x)\partial_x\overline{\rho}_{\mathrm{s}}(x,V)+2\pi \sigma\frac{eV}{k_{\mathrm{B}}T}\frac{R_{\mathrm{t}}R_{\mathrm{b}}}{R(x)L}\right)\\
			&+Q(V)\overline{\rho}_{\mathrm{s}}(x,V)
		\end{split}\tag{S3}
	\end{equation}
	which represents the diffusive, conductive and advective components, respectively and where $Q(V)=-\frac{V}{L}\pi R_{\mathrm{t}}R_{\mathrm{b}}\frac{\epsilon\psi_0}{\eta}$ is the electro-osmotic volume flow. The net total number of ions entering the channels is given by the difference in salt flux between the tip and base $J_x(L)-J_x(0)$. Since we consider a small voltage perturbation $V^{\prime}$ around $V=0$, we consider an initially homogeneous state $\bar{\rho}_{\rm{s}}(x)=2\rho_{\rm{b}}$ where $\partial_x\overline{\rho}_{\mathrm{s}}(x,V)=0$. In this case the first term in Eq.~(\ref{eq:saltflux}) vanishes and also the third term vanishes as $Q$ is laterally constant and $\overline{\rho}_{\mathrm{s}}(0,V)=\overline{\rho}_{\mathrm{s}}(L,V)$. The resulting net salt flux at a small static potential $V^{\prime}$ is then purely given by the conductive terms
	\begin{align}\label{eq:netflux}
		J_x(0)-J_x(L)=2\pi\frac{D\Delta R}{L}\frac{e\sigma}{k_{\mathrm{B}}T}V^{\prime}\equiv \gamma V^{\prime},\tag{S4}
	\end{align}
	where $\gamma<0$ for our parameter choice of $\sigma<0$ and $\Delta R>0$. With $\alpha$ and $\gamma$ defined in Eqs.~(\ref{eq:alpha}) and (\ref{eq:netflux}), respectively, we obtain the timescale $\tau$ of interest as
	\begin{equation}\label{eq:tsSM}
		\tau=\frac{\alpha}{\gamma}= \frac{1}{12}\frac{L^2}{D},\tag{S5}
	\end{equation}
	via which we see that the timescale is dictated by the characteristic diffusion time of the channel. This is a surprising result as the $\gamma V^{\prime}$ of Eq.~\ref{eq:saltflux} term is purely conductive. Eq.~(\ref{eq:tsSM}) is plotted in the right inset of Fig.~2, where we compare $\tau$ with the reciprocal frequencies $1/f_{\mathrm{max}}$ from full finite-element for which the enclosed area in the I-V hysteresis loop is maximal for various lengths, for all $L$ they are related via $f_{\mathrm{max}}\tau\approx 0.19$.
	
	To arrive at Eq.~(\ref{eq:netflux}), through which Eq.~(\ref{eq:rhosProf}) is obtained in Ref.~\cite{boon2021nonlinear}, the assumption is made that $\lambda_{\mathrm{D}}\ll R(x)$, which we mildly violate around the tip of the channel for our standard parameter set. Nevertheless, we still obtain good agreement with finite element calculations, however we note that our analytical approximation is not necessarily universally applicable to all parameter sets. We remark that in our work the emergence of neuromorphic behaviour was rather sensitive to changes in the parameters. Although we obtained spiking for numerous different parameter sets, this sensitivity suggests that explorations of the parameter space would be required in experiments.
	
	\section{Explanation of spike train and frequency modulation}
	Eqs.~(7) and (9) form a dynamical system of equations, which we can analyse in some more detail to gain a deeper understanding of the spiking behaviour presented in Fig.~3. Firstly, since $\tau_{\pm}\ll\tau_{\mathrm{s}}$ we can assume instantaneity of the $g_{\pm}$ channels, i.e.\ $g_{\pm}(V(t),t)=g_{\infty,\pm}(V(t))$. With this assumption the current contribution in the circuit in Fig.~3(b) through the fast channels is determined directly by Eq.~(5), i.e.\ 
	\begin{equation}\label{eq:F}
		\begin{split}
			g_{\mathrm{r}}F(V_{\mathrm{m}}(t))\equiv&-g_{\infty,+}(-V_{\mathrm{m}}(t)+E_{+})\left(V_{\mathrm{m}}(t)-E_{+}\right)\\
			&-g_{\infty,-}(V_{\mathrm{m}}(t)-E_{-})\left(V_{\mathrm{m}}(t)-E_{-}\right),
		\end{split}\tag{S6}
	\end{equation}
	representing the total current from the fast channels and which we denote by $g_{\mathrm{r}}F(V_{\mathrm{m}}(t))$ where $g_{\mathrm{r}}$ is a characteristic effective conductance of the fast channels. By expanding Eq.~(\ref{eq:F}) in $V_{\mathrm{m}}(t)$, we find that $F(V_{\mathrm{m}}(t))$ is well-approximated by
	\begin{align}\label{eq:Fappr}
		g_{\mathrm{r}}F(V_{\mathrm{m}}(t))\approx\mathcal{G}V_{\mathrm{m}}(t)-\frac{(\mathcal{G}V_{\mathrm{m}}(t))^3}{3V_{\mathrm{r}}^2},\tag{S7}
	\end{align}
	where the characteristic effective conductance of the fast channels $g_{\mathrm{r}}=1\;\mathrm{pS}$, a reference voltage $V_{\mathrm{r}}=1\;\mathrm{V}$ and a dimensionless parameter $\mathcal{G}=3.5$ are determined by expanding $F(V_{\mathrm{m}}(t))$ around $V_{\mathrm{m}}(t)=0$.
	
	This assumption reduces the dynamical system to a set of only two ordinary differential equations, given by
	\begin{align}
		\tau_{\mathrm{m}}\dfrac{\dd V_{\mathrm{m}}(t)}{\dd t}=&\frac{I(t)}{g_{\mathrm{r}}}-\frac{g_{\mathrm{s}}(t)}{g_{\mathrm{r}}}\left(V_{\mathrm{m}}(t)-E_{\mathrm{s}}\right)+F(V_{\mathrm{m}}(t)),\tag{S8}\label{eq:2DV}\\
		\tau_{\mathrm{s}}\dfrac{\dd g_{\mathrm{s}}(t)}{\dd t}=& g_{\infty,\mathrm{s}}(-V_{\mathrm{m}}(t)+E_{\mathrm{s}})-g_{\mathrm{s}}(t),\tag{S9}\label{eq:2Dg}
	\end{align}
	where we defined a membrane response RC time of $\tau_{\mathrm{m}}=C/g_{\mathrm{r}}=5$ ms, for the parameter set used in Fig.~3. With Eqs.~(\ref{eq:2DV}) and (\ref{eq:2Dg}) we obtain results that are essentially indistinguishable from those in Fig.~3.
	
\begin{figure}[ht]
		\centering
		\includegraphics[width=0.46\textwidth]{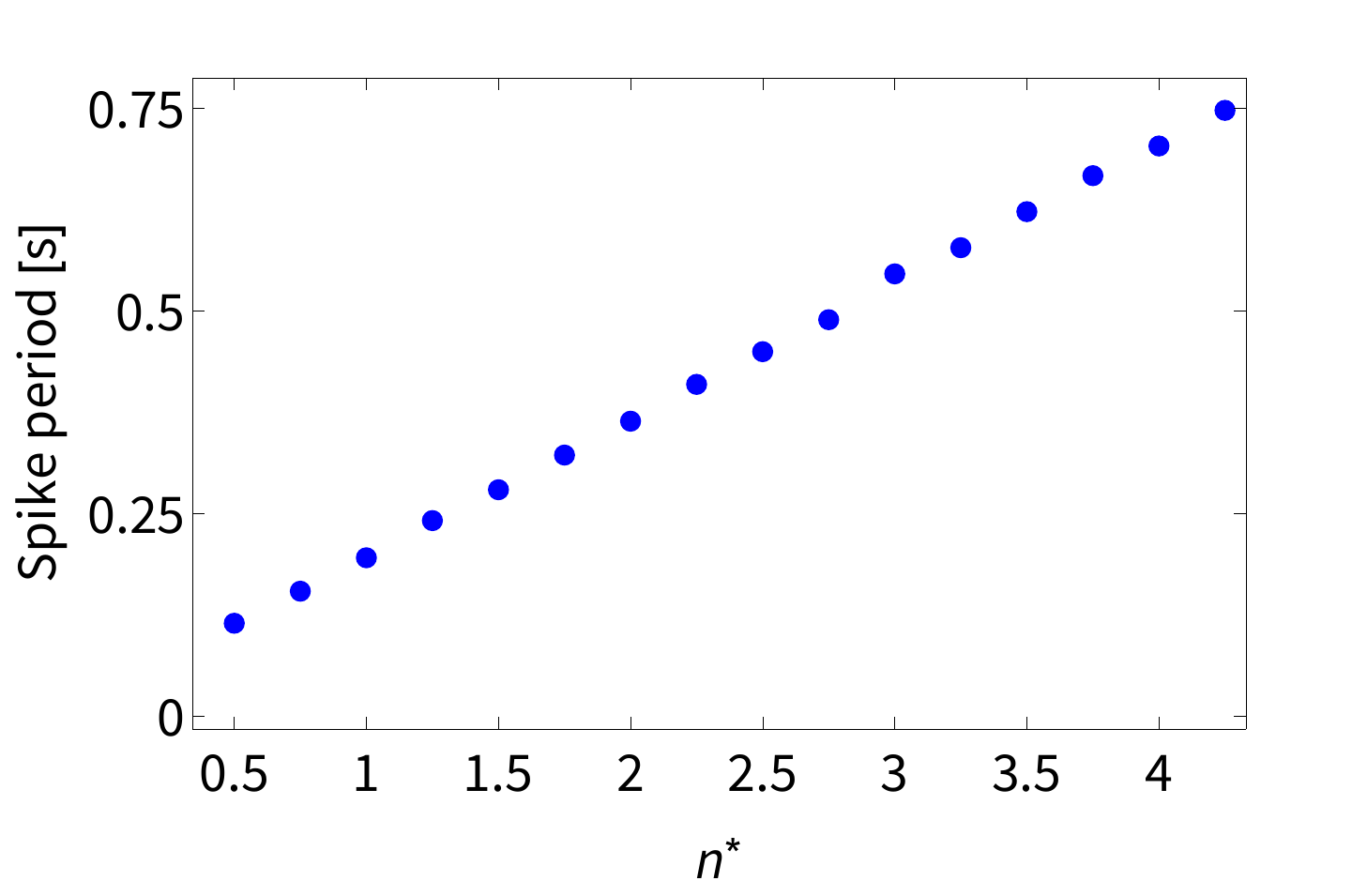}
		\caption{The spike train period from the circuit presented in Fig.~3(b) as a result of the full system of Eqs.~(7) and (9) when the timescales $\tau_{\ch{m}}$ and $\tau_{\ch{s}}$ are rescaled by a factor $ n^{*}$, i.e.\ $\tau_{\ch{m}}\to n^{*}\tau_{\ch{m}}$ and $\tau_{\ch{s}}\to n^{*}\tau_{\ch{s}}$.}
		\label{fig:spikefreq}
	\end{figure}

	Eqs.~(\ref{eq:2DV}) and (\ref{eq:2Dg}) suggest that it is possible to modulate the spike train frequency by altering the two timescales $\tau_{\mathrm{m}}$ and $\tau_{\mathrm{s}}$. Physically this could be achieved by tuning the capacitance $C$ and the slow channel length $L_{\ch{s}}$, respectively. If we scale both $\tau_{\ch{m}}$ and $\tau_{\ch{s}}$ by the same factor $ n^{*}$, i.e.\ $\tau_{\ch{m}}\to n^{*}\tau_{\ch{m}}$ and $\tau_{\ch{s}}\to n^{*}\tau_{\ch{s}}$, we indeed see in Fig.~\ref{fig:spikefreq} that the spike train period depends essentially linearly on this factor $ n^{*}$, where we again solved the full system of Eqs.~(7) and (9).
	
	We observe that Eqs.~(\ref{eq:2DV}) and (\ref{eq:2Dg}) are very similar to the FitzHugh-Nagumo (FN) model (also known as the Bonhoeffer-van der Pol model), which is of no surprise since these models serve as simplified versions of the HH model \cite{fitzhugh1961impulses,nagumo1962active}. In particular, Eqs.~(\ref{eq:2DV}) and (\ref{eq:2Dg}) are formulated similar to the system investigated in Ref.~\cite{bisquert2021frequency}. Although the FN model is not identical to Eqs.~(\ref{eq:2DV}) and (\ref{eq:2Dg}), we do believe that both the all-or-none behaviour and the spike train behaviour can be understood through the FN model, an approach that was used to understand the original HH model results as well \cite{fitzhugh1961impulses,nagumo1962active,troy1976bifurcation,rabinovitch1999threshold,bisquert2021frequency}.
	
	A mathematical understanding of the spiking behaviour through the FN model has been broadly investigated \cite{fitzhugh1961impulses,nagumo1962active,troy1976bifurcation,rabinovitch1999threshold,bisquert2021frequency}. Additionally we offer here a heuristic explanation of why we see emerging neuromorphic behaviour from our specific iontronic circuit. If we consider the circuit in Fig.~3(b) without the slow channel and with $I(t)=0$, then we find two stable stationary points at $V_{\mathrm{m}}\approx\pm0.5\text{ V}$.  These stationary points correspond to the non-trivial roots of $F(V_\mathrm{m})$. This bistability is a precursor of the all-or-none law we find. 
	\begin{figure}[ht]
		\centering
		\includegraphics[width=0.46\textwidth]{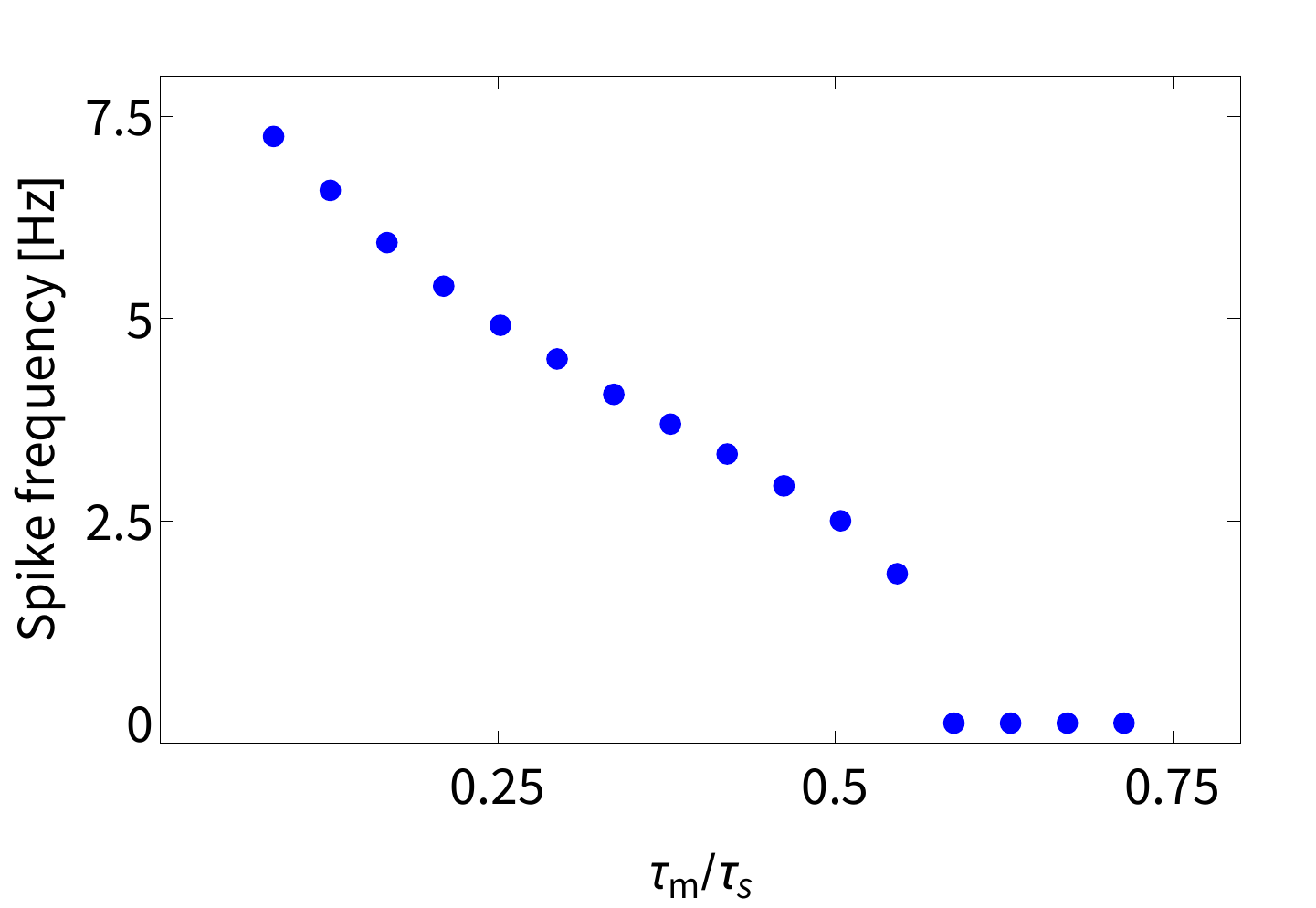}
		\caption{The spike train frequency from the circuit presented in Fig.~3(b) as a result of the full system of Eqs.~(7) and (9) when the timescales $\tau_{\ch{m}}$ is changed while $\tau_{\ch{s}}$ is kept constant. The spike train only emerges if $\tau_{\ch{m}}<0.55\tau_{\ch{s}}$.}
		\label{fig:spikefreq2}
	\end{figure}
	For a sufficiently strong imposed stimulus current $I(t)=I>0$, the negative stationary state vanishes and only a positive stable stationary state $V_{\mathrm{m}}>0$ remains. The $V_{\mathrm{m}}<0$ state for $I(t)=0$ also emerges in the full circuit with the slow channel (Fig.~3(b)), which we observe as the initial resting state as seen in Figs.\ 3(d) and 3(f). A $V_{\mathrm{m}}>0$ state is not present here because the slow channel becomes more conductive for $V_{\mathrm{m}}>0$ and pushes $V_{\mathrm{m}}(t)$ to the battery potential $E_{\mathrm{s}}<0$. Now assume the full system is at rest with $I(t)=0$ at $V_{\mathrm{m}}(t)=V_{\mathrm{m}}^*<0$ and we increase $I(t)$ at time $t^{\prime}$ from $0$ to some $I>I_{\mathrm{AP}}$. Then after some time $\Delta t\ll \tau_{\mathrm{s}}$, such that $g_{\mathrm{s}}(t^{\prime}+\Delta t)\approx g_{\mathrm{s}}(t^{\prime})=g_{\infty,s}(V_{\mathrm{m}}^*)$, the slow channel is still in a low conducting state. If $V_{\mathrm{m}}(t)$ is able to significantly change over a time period $\Delta t$, then the system can transition to the $V_{\mathrm{m}}>0$ stationary state before the slow channel becomes conductive enough to make this state vanish. Note that this requires that the RC-like membrane voltage response time $\tau_{\ch{m}}$ is much shorter that the slow channel timescale, i.e.\ $\tau_{\mathrm{m}}\ll\tau_{\mathrm{s}}$.  In Fig.~\ref{fig:spikefreq2}, which shows the spike train frequency as a function of $\tau_{\mathrm{m}}/\tau_{\mathrm{s}}$, we show that this requirement actually also holds for the full circuit presented in Fig.~3(b). A spike train only emerges when $\tau_{\ch{m}}<0.55\tau_{\ch{s}}$ and for $\tau_{\ch{m}}>0.55\tau_{\ch{s}}$ the spike train does not emerge, corresponding to a frequency of 0 in Fig.~\ref{fig:spikefreq2}. This fast-slow relation is characteristic for FN models \cite{fitzhugh1961impulses} and this requirement is also found in Ref.~\cite{bisquert2021frequency}.
	
	The all-or-none behaviour can now be explained by the observation that the $V_{\mathrm{m}}<0$ stationary state either remains in place for $I<I_{\mathrm{AP}}$ (thus no action potential) or vanishes temporarily for $I>I_{\mathrm{AP}}$, resulting in a complete action potential. For a stimulus current $I_{\ch{AP}}<I<I_{\mathrm{train}}\approx 1.28\text{ pA }$ there is still a stable stationary point, which can be seen in the subcritical voltage trace in Fig.~3(f, red). At $I=I_{\mathrm{train}}\approx 1.28\text{ pA }$, this stable point undergoes a Hopf bifurcation \cite{hale2012dynamics} and the trace shows a periodic solution, as shown in Fig.~3(f, blue). After the Hopf bifurcation two more stationary points appear inside the stable limit cycle which are both also unstable, thus it is a supercritical Hopf bifurcation \cite{lynch2004dynamical}. 
	
\begin{figure}[ht]
		\centering
		\includegraphics[width=0.46\textwidth]{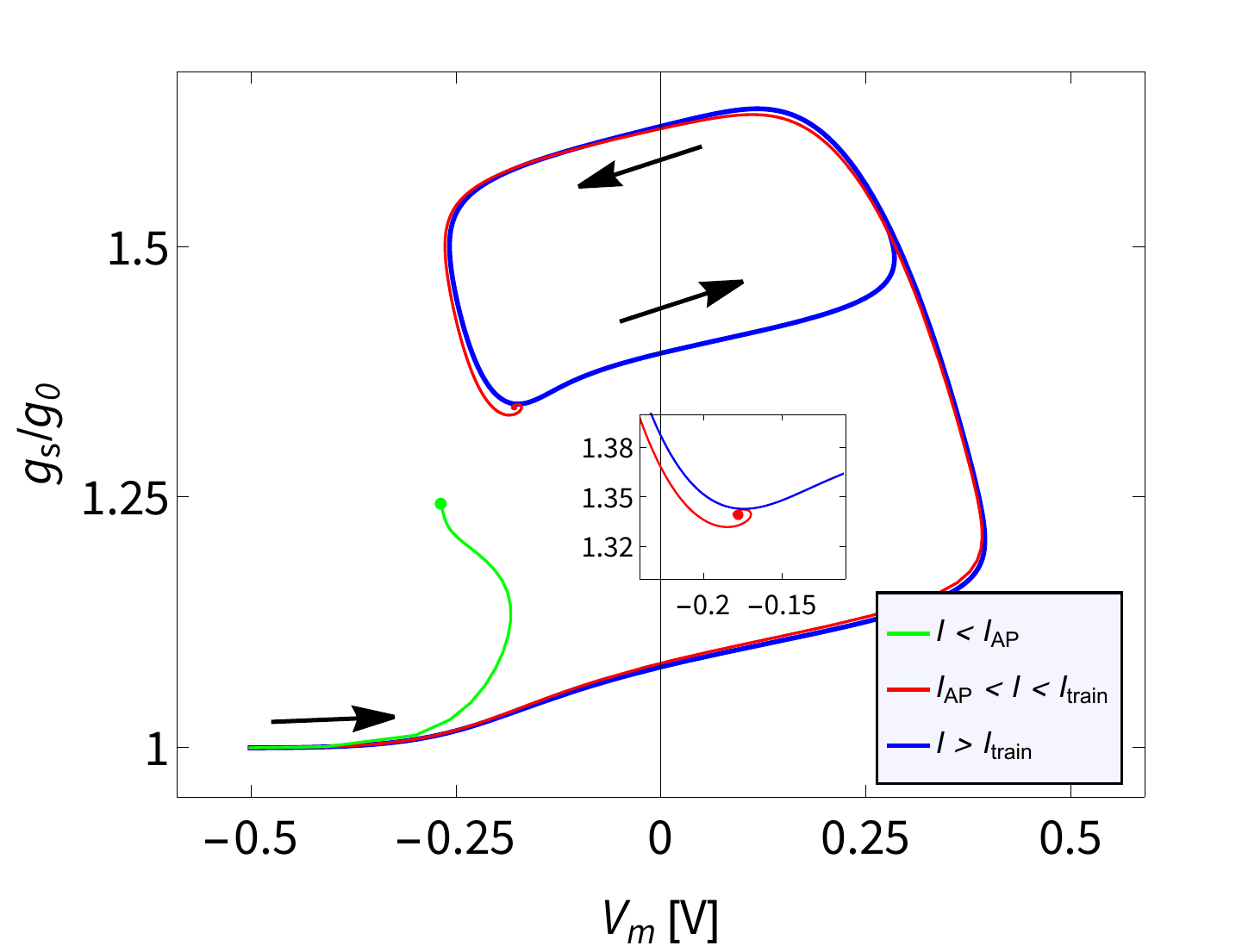}
		\caption{The phase portrait of $V_{\ch{m}}(t)$ and $g_{\ch{s}}(t)$, obtained by solving the two-dimensional set of equations (\ref{eq:2DV}) and (\ref{eq:2Dg}). The trajectories of $V_{\ch{m}}(t)$ and $g_{\ch{s}}(t)$ start out from a resting state with $I=0$, after which sustained currents of $I=1.16\text{ pA }<I_{\ch{AP}}$, $I_{\ch{AP}}<I=1.27\text{ pA }<I_{\ch{train}}$ and $I=1.28\text{ pA }>I_{\ch{train}}$ resulting in the green, red and blue trajectories, respectively.}
		\label{fig:phaseplot}
	\end{figure}

	With the identification that the relevant dynamical variables are $V_{\ch{m}}(t)$ and $g_{\ch{s}}(t)$ we can alternatively display the results from Figs.\ 3(d) and 3(f) in a phase portrait of $V_{\ch{m}}(t)$ and $g_{\ch{s}}(t)$. In Fig.~\ref{fig:phaseplot} we show the trajectories of $V_{\ch{m}}(t)$ and $g_{\ch{s}}(t)$ which start out from a resting state with $I=0$, after which sustained currents of $I=1.16\text{ pA }<I_{\ch{AP}}$, $I_{\ch{AP}}<I=1.27\text{ pA }<I_{\ch{train}}$ and $I=1.28\text{ pA }>I_{\ch{train}}$ are imposed, resulting in the green, red and blue trajectories, respectively. The green trajectory settles to a new stationary state rather directly, while the red trajectory first traverses a single orbit through the $\left(V_{\ch{m}}(t),g_{\ch{s}}(t)\right)$ space, which is visible as the single action potential shown in Fig.~3(d). The supercritical Hopf bifurcation at $I=1.28\text{ pA}$ translates to the blue periodic orbit, corresponding to the spike train shown in Fig.~3(f).
	
	%
